\documentclass[prb,final,showpacs,onecolumn,superscriptaddress,aps]{revtex4-2}
\usepackage[margin=3cm]{geometry}
\usepackage[utf8]{inputenc}
\usepackage[american,british]{babel}
\usepackage[T1]{fontenc}
\usepackage[pdftex]{graphicx}  
\usepackage{graphicx, xcolor}
\usepackage{dcolumn}
\usepackage{braket}
\usepackage{bm}
\usepackage{amsmath,amsthm,amssymb}
\usepackage{color}
\usepackage{verbatim}
\usepackage{lipsum}

\usepackage[colorlinks,linkcolor=blue,citecolor=blue,urlcolor=blue]{hyperref}

\usepackage{dsfont}
\usepackage{physics}

\usepackage{soul}

\newcommand{\me}[1]{\left\langle #1 \right\rangle }

\begin{document}

\title{Villain model with long-range couplings}

\author{Guido Giachetti}
\email{ggiachet@sissa.it}
\affiliation{SISSA and INFN Sezione di Trieste, Via Bonomea 265, I-34136 Trieste, Italy}

\author{Nicol\`o Defenu}
\affiliation{Institut f\"ur Theoretische Physik, ETH Z\"urich, Wolfgang-Pauli-Str.\,27 Z\"urich, Switzerland}

\author{Stefano Ruffo}
\affiliation{SISSA and INFN Sezione di Trieste, Via Bonomea 265, I-34136 Trieste, Italy}
\affiliation{Istituto dei Sistemi Complessi, Consiglio Nazionale delle Ricerche, Via Madonna del Piano 10, I-50019 Sesto Fiorentino, Italy}

\author{Andrea Trombettoni}
\affiliation{Department of Physics, University of Trieste, Strada Costiera 11, I-34151 Trieste, Italy}
\affiliation{SISSA and INFN Sezione di Trieste, Via Bonomea 265, I-34136 Trieste, Italy}

\begin{abstract}
\noindent
The nearest-neighbor Villain, or periodic Gaussian, model is a useful tool to understand the physics of the topological defects of the two-dimensional nearest-neighbor $XY$ model, as the two models share the same symmetries and are in the same universality class. The long-range counterpart of the two-dimensional $XY$ has been recently 
shown to exhibit a non-trivial critical behavior, with a complex phase diagram including a range
of values of the power-law exponent of the couplings decay, $\sigma$, in which there are a magnetized, a disordered and a critical phase\,\cite{Giachetti2021}.
Here we address the issue of whether the 
critical behavior of the two-dimensional $XY$ model with long-range couplings
can be described by the Villain counterpart of the model. After introducing a suitable generalization of the Villain model with long-range couplings, we derive a set of renormalization-group equations for the vortex-vortex potential, which differs from the one of the long-range $XY$ model, signaling that the decoupling of spin-waves and topological defects is no longer justified in this regime.
The main results are that for $\sigma<2$ the two models no longer share the same universality class. Remarkably, within a large region of its the phase diagram, the Villain model is found to behave similarly to the one-dimensional Ising model with $1/r^2$ interactions.  
\end{abstract}

\maketitle

\section{Introduction}
\noindent
In the context of the study of critical phenomena, the universality class of the two-dimensional (or $1+1$) $XY$ model holds a special place, as its phenomenology is radically different from those of the other $O(n)$ models. 
As both bosonic and fermionic systems described by complex order parameters naturally exhibit a $U(1)$ symmetry, such an universality class is important to describe the physics of a large variety of high energy, condensed matter and field theory systems, including superconductors,
Helium, spin models, and the complex $|\phi|^4$ theory \cite{goldenfeld2018lectures}. A remarkable property is that, even in absence of a  
finite order parameter, the model undergoes a phase transition -- usually referrend to as Berezinskii-Kosterlitz-Thouless (BKT) -- between a phase with power-law correlation functions, described in the infrared by a line of fixed points, and a disordered one\,\cite{KT,kosterlitz2017nobel,savit}. The study of the nearest-neighbor $XY$ model is made easier by the introduction of the so-called Villain model, introduced in \cite{Villain1974}. i.e. a simplified version of the $XY$ Hamiltonian, which is able to faithfully reproduce the features of the BKT transition in two dimensions\,\cite{jose1977renormalization, kleinert1989gauge}. The advantage of the Villain model is that it allows for an exact mapping to the Hamiltonian of a Coulomb gas, in which topological defects interact through a logarithmic potential as charged particles\,\cite{savit}. Within this Coulomb gas picture, the BKT transition can be understood in terms of the unbinding of vortex pairs. While this mapping is no longer present in higher dimensions, further studies suggests that, also in this case, the two models share the same universality class \cite{kleinert1989gauge,dario2020massless}.

Beyond its link to the $XY$ model, the Villain Hamiltonian has drown considerable attention \emph{per se}, proving interesting both from the theoretical\,\cite{Frolich1981,Kapikranian2008,Haraldsen2009,Aizenman2021,gabay1989phase,Gorantla2021} and numerical\,\cite{janke1986good,Janke1993,Janke1997,Hasenbusch2005,surungan2019berezinskii} point of view, with applications running from the study of quantum-phase transitions\,\cite{witczak2014dynamics} and superconductivity\,\cite{dasgupta1981phase} to lattice gauge theories\,\cite{kleinert1989gauge,onofri1981,Romo2012, dario2020massless, Choi2022noninvertible}, deconfinement\,\cite{borisenko2015deconfinement} and duality\,\cite{palan1987duality} in high-energy physics. It is thus customary in the literature to refer to the
{\it Villain approximation} when the coupling is adjusted to reproduce in the best way the $XY$ model free energy,
and to the {\it Villain model} when the model is studied by itself\,\cite{kleinert1989gauge}. The fate of the model in presence of long-range couplings, which is the subject of this work, is thus an interesting problem in itself. 


The addition of non-local, long-range potentials between the microscopic components of the system, is known to give rise to plenty of new physics of both classical\,\cite{campa2014physics} and quantum\,\cite{defenu2021longrange} many-body systems. The study of this models have recently sparked a new wave of interest, due \textcolor{black}{to} the possibility of experimental realizations in atomic, molecular
and optical (AMO) systems\,\cite{haffner2008quantum,lahaye2009physics,saffman2010quantum, ritsch2013cold,bernien2017quantum,monroe2021programmable,defenu2021longrange,mivehvar2021cavity}. In particular, as the hypotheses of the celebrated Hohenberg-Mermin-Wagner theorem\,\cite{MW} are no longer met, long-range interactions can induce a 
spontaneous symmetry breaking (SSB) in a low-dimensional system, as discussed in the classic papers\,\cite{Dyson1969,Thouless1969,Kosterlitz76,cardy1981one}. In any case, sufficiently slow-decaying interactions are known to alter the universal features of the critical behavior of the systems, e.g. the critical exponents\,\cite{Kosterlitz76, defenu2015fixed}. 

For the case of the classical $O(n)$ model, the relevance of long-range interactions to the critical behavior can be understood through the Sak criterion\,\cite{sak1973recursion}. If we consider power-law couplings of the form $\sim 1/r^{d+\sigma}$, where $d$ is the system dimension, the criterion states that the addition of long-range couplings can affect the critical properties of the system as long as $\sigma < \sigma^{*} = 2 - \eta_{\rm sr}$, $\eta_{\rm sr}$ being the anomalous dimension in the nearest-neighbours limit ($\sigma \rightarrow \infty$). For $\sigma > \sigma^{*}$ we recover the short-range regime. 

While the criterion 
has been investigated carefully\,\cite{Luijten,Defenu2020review}, the $n=2$, $d=2$ case (i.e. the two-dimensional $XY$ model) anyway escapes to such criterion. As already mentioned, in this case the critical behavior of the short-range counterpart of the model is peculiar, as its low-temperature phase 
is not described, at the renormalization-group (RG) level, by an isolated fixed point, but rather by whole line of fixed points. As a consequence the short-range anomalous dimension $\eta_{\rm sr}$ is not unambiguously defined. 

It has been recently shown\,\cite{Giachetti2021,Giachetti2022prb,Giachetti2021EPL} that the addition of long-range couplings to the two-dimensional $XY$ model gives rise to a complex phase diagram, showing that the transition between the short-range regime and the long-range one cannot be captured within Sak's picture. It is however unclear whether, in the long-range regime, $XY$ model still allows for such description in terms of topological charges, and whether a generalization of the Villain model to long-range interactions is able to reproduce the complex critical behavior of the long-range $XY$ model. Loosely speaking, long-range interactions are known to effectively increase the dimensionality of the systems (see \cite{angelini2014relations,defenu2015fixed} and Refs. therein); while this would suggest that the two models exhibit the same critical behavior even in the long-range regime, we are going to see that the case of the two-dimensional Villain model is considerably subtler, due to the interplay between topological excitations and long-range couplings.


The paper is structured as follows. After discussing 
a way to generalize the Villain model to the case of long-range couplings in Sec. \ref{SecDefVIll}, in Sec. \ref{SecVortexVortex} we derive the vortex-gas description of the model, analogous to the Coulomb gas one. In Sec. \ref{SecRealRG} we derive the corresponding real-space RG equations and we describe the corresponding phase diagram. Finally, in Sec. \ref{SecQFTVill} we comment on our result in the light of 
field-theoretical description of the model.

\section{Definition of the model}
\label{SecDefVIll}
\noindent
We consider thus a set of $N$ planar spins $\mathbf{s}_{\mathbf{j}}$, such that $\mathbf{s}_{\mathbf{j}}^2 = 1$, arranged on a two-dimensional square lattice. If we parameterize each spin with the phase $\theta_{\mathbf{j}}$ as $\mathbf{s}_{\mathbf{j}} \equiv \left(\cos \theta_{\mathbf{j}}, \sin \theta_{\mathbf{j}} \right)$, the $XY$ Hamiltonian, for a generic choice of the couplings, takes the form
\begin{equation}\label{model_xy}
\beta H = \frac{1}{2} \sum_{\mathbf{i} \neq \mathbf{j}} J(r) \left[1 - \cos ( \theta_{\mathbf{i}} - \theta_{\mathbf{j}})\right]
\end{equation}
with $\mathbf{i}, \mathbf{j} \in \mathbb{Z}^2$, $\mathbf{r} = \mathbf{i}-\mathbf{j}$, $r = |\mathbf{r}|$. The model exhibits a global $O(2)$ symmetry which, in terms of the angular variables $\theta_{\mathbf{j}}$, can be written as $\theta_{\mathbf{j}} \rightarrow \theta_{\mathbf{j}} + \alpha$. On top of this, the phases $\theta_{\mathbf{j}}$ are periodic, which results into the local symmetry
\begin{equation} \label{localsym}
    \theta_{\mathbf{j}} = \theta_{\mathbf{j}} + 2 \pi N_{\mathbf{j}}
\end{equation}
where, for each $\mathbf{j}$, $N_{\mathbf{j}} \in \mathbb{Z}$. This symmetry is indeed crucial, in the short-range case, to allow for the presence of vortices, i.e. topological configurations in which the variable $\theta_{\mathbf{j}}$ increases of an integer multiple of $2 \pi$ as we follow a closed loop on the lattice. 

In the nearest-neighbor case, the Hamiltonian of the Villain model is instead given by
\begin{equation} \label{SRVillH}
    H = \frac{J}{2} \sum_{\me{\mathbf{i},\mathbf{j}}} \left( \theta_{\mathbf{i}} - \theta_{\mathbf{j}} - 2 \pi n_{\mathbf{i},\mathbf{j}} \right)^2,
\end{equation}
where $\mathbf{i}, \mathbf{j} \in \mathbb{Z}^2$, the $\theta_{\mathbf{j}} \in \mathbb{R}$  while $n_{\mathbf{i},\mathbf{j}} \in \mathbb{Z}$ are discrete link variables which couples nearest-neighbor pairs and obey the relation $n_{\mathbf{i},\mathbf{j}} = - n_{\mathbf{j},\mathbf{i}} $.

The main advantage of the model is that the presence of this auxiliary integer link variables is able to reproduce the periodicity of the interaction in the $XY$ Hamiltonian \eqref{model_xy} (with $J(r) = J \delta_{r,1}$), without introducing a direct interaction between the angular variables. This can be seen explicitly by noticing that Hamiltonian \eqref{SRVillH} is invariant under the local transformation
\begin{equation} \label{gaugesym}
\begin{split}
    \theta_{\mathbf{j}} &\rightarrow \theta_{\mathbf{j}} + 2 \pi N_{\mathbf{j}} \\
    n_{\mathbf{i},\mathbf{j}} &\rightarrow N_{\mathbf{j}} - N_{\mathbf{i}}, 
\end{split}
\end{equation}
with $N_{\mathbf{i}} \in \mathbb{Z}$, which can be thought of as a discrete gauge symmetry on the lattice \cite{savit,kleinert1989gauge}. Restricted to the continuous variables $\theta_{\mathbf{j}}$ this is exactly the local symmetry \eqref{localsym} of the original $XY$ model, so that the model can account for its non-trivial critical properties due to the presence of topological configurations. 

Now we want to introduce the generalization of the Villain model to the case of long-range, power-law decaying couplings, i.e. 
\begin{equation} \label{Jr}
     J(\mathbf{r}) \sim J {r^{-2-\sigma}}
\end{equation}  
for $r \gg 1$. Here we choose $\sigma > 0$ in order to preserve the extensive nature of the thermodynamical quantities \cite{kac1963van, campa2014physics}. The naive generalization of the Hamiltonian \eqref{SRVillH}, in which the sum runs all over the lattice sites $\mathbf{i}$,$\mathbf{j}$ and $J$ is replaced by Eq.\, \eqref{Jr}, is problematic, as in this case we would have to deal with link-variables whose number grows super-extensively (as $O(N^2)$). 

In order to overcome this problem, and to define a sensitive generalization of the model, we have to give a more general definition of the Villain model. For our proposes, thus, we define the Villain model as a quadratic model in the $\theta_{\mathbf{j}}$, which preserves both the global $O(2)$ and the local symmetry of Eq.\,\eqref{localsym}. This more general definition, not only allows us to deal with a generic choice of the couplings, but also defines naturally the continuum limit of the model in a field-theoretical language. 
We observe that alternative choices could possibly be made, and although one should certainly investigate whether fall or not into the same universality class, one can anyway expect
that this is indeed the case.

We now see how to explicitly construct the Villain Hamiltonian corresponding to an $XY$ model with a generic choice of the couplings $J(r)$. To better understand the procedure, however, it is more convenient to address firstly the continuous version of the $XY$ Hamiltonian, in which $\theta_{\mathbf{j}}$ is replaced by a continuous field $\theta(\mathbf{x})$. The so-called Berezinskii approximation, i.e. the substitution
\begin{equation} \label{quadraticapprox}
    1-\cos\left( \theta (\mathbf{x+r})-\theta (\mathbf{x}) \right) \rightarrow \frac{1}{2} \left( \theta (\mathbf{x+r})-\theta (\mathbf{x}) \right)^2, 
\end{equation}
breaks the local symmetry \eqref{localsym}, which accounts for the periodicity of $\theta(\mathbf{x})$. In the continuum limit, this property can be stated by saying that $\theta (\mathbf{x})$ is not a single-valued function (but rather is defined up to integer multiples of $2 \pi$) so that for each closed path $\partial A$ on the plane
\begin{equation} \label{topchargecont}
    \oint_{\partial A} \nabla \theta (\mathbf{x}) \cdot d \mathbf{x} =  \int_{A} \nabla \times \nabla \theta (\mathbf{x}) \ d^2 \mathbf{x} = 2 \pi M(A)
\end{equation}
with $M(A) \in \mathbb{Z}$ and $\nabla \times \mathbf{a} = \epsilon_{j,k} \partial_j a_k$, $\epsilon_{j,k}$ being the completely antisymmetric rank-$2$ tensor. If we divide the region of the plane $A$ enclosed by $\partial A$ into two subregions $A_1$, $A_2$ we will have that $M(A) = M(A_1) + M(A_2)$ so that $M(A)$ has the meaning of the total topological charge enclosed into the region of the plane $A$. Accordingly, Eq.\,\eqref{topchargecont} can also be put in a local form, namely 
\begin{equation} \label{constraintvort}
    \nabla \times \nabla \theta (\mathbf{x}) = 2 \pi \sum_{k} m_k \delta( \mathbf{x} - \mathbf{x}_k),
\end{equation}
where $m_k \in \mathbb{Z}$ and $\mathbf{x}_k$ can be interpreted as vortex charges and thepositions respectively. In the continuum limit then the Villain model can be defined as a boson $\theta(\mathbf{x})$, interacting with point-like charges $m_k$ through the constraint \eqref{constraintvort}. In formal terms 
\begin{equation} \label{ZVillcont}
Z = \sum_{\lbrace m_i \rbrace} \int \mathcal{D}(\nabla \theta) \ e^{-\beta H_0 (\theta)} \delta( \nabla \times \nabla \theta  - 2 \pi n (\mathbf{x})),
\end{equation}
where the sum represent the trace all over all the possible vortex configurations, $\beta H_0$ the quadratic Hamiltonian
\begin{equation} 
\beta H_0 = \frac{1}{4} \int d^2 \mathbf{x} d^2 \mathbf{r} \ J(r) \left( \theta (\mathbf{x+r})-\theta (\mathbf{x}) \right)^2, 
\end{equation}
and we introduced the vortex density
\begin{equation} \label{vortexdensity}
n (\mathbf{x})= \sum_{k} m_k \delta( \mathbf{x} - \mathbf{x}_k). 
\end{equation}
Let us notice how we are now integrating over the configurations of $\nabla \theta$, which are single-valued, instead of $\theta(\mathbf{x})$.

Let us now see how we can built the Villain model directly on lattice. Once again, the quadratic approximation alone, namely
\begin{equation}
    1-\cos(\theta_{\mathbf{j+r}}-\theta_\mathbf{j}) \rightarrow \frac{1}{2} (\theta_{\mathbf{j+r}}-\theta_\mathbf{j})^2.
\end{equation}
would break the local symmetry of Eq.\,\eqref{localsym}, and thus the possibility of correctly describe topological configuration. Indeed, given a closed loop of $P$ points on the lattice $\mathbf{j}_1$,$\mathbf{j}_2$,$\dots$,$\mathbf{j}_P$,$\mathbf{j}_{P+1} \equiv \mathbf{j}_{1}$ the lattice equivalent of the integral in Eq.\eqref{topchargecont} is given by 
\begin{equation}
    \sum^{P}_{p=1} \left(\theta_{\mathbf{j}_{p+1}}- \theta_{\mathbf{j}_p} \right) \equiv 0 .
\end{equation}
In order to overcome this problem,  we follow the original idea of the seminal Villain's work i.e. we introduce an integer-valued link variable $n_{\mathbf{i}, \mathbf{j}}$ ($n_{\mathbf{i}, \mathbf{j}} = - n_{\mathbf{j}, \mathbf{i}}$) for each paof lattice points  and we make the further replacement 
\begin{equation} \label{villainlink}
    \theta_{\mathbf{j+r}}-\theta_\mathbf{j} \rightarrow \theta_{\mathbf{j+r}}-\theta_\mathbf{j} + 2 \pi n_{\mathbf{i}, \mathbf{j}} . 
\end{equation}
As a consequence, the Villain Hamiltonian becomes
\begin{equation} \label{VHam}
\beta H_0 = \frac{1}{4} \sum_{\mathbf{i} \neq \mathbf{j}} J(r) \left( \theta_{\mathbf{i}}-\theta_\mathbf{j} - 2 \pi n_{\mathbf{i},\mathbf{j}} \right)^2, \end{equation}
(with $r = |\mathbf{j}-\mathbf{i}|$) while the closed-loop integral of Eq.\eqref{topchargecont} is now
\begin{equation}
    \sum^{P}_{p=1} \left(\theta_{\mathbf{i}_{p+1}}- \theta_{\mathbf{j}_p} - 2 \pi n_{\mathbf{j}_p,\mathbf{j}_{p+1}} \right) = 2 \pi \sum^{P}_{p=1} n_{\mathbf{j}_p,\mathbf{j}_{p+1}} . 
\end{equation}
In this language, the lattice-analogous of the constraint Eq.\eqref{topchargecont} imposes that the charge enclosed in the region $A$ defined by the $\mathbf{j}_p$, \begin{equation}
    M (A) = \sum_{\partial A} n_{\mathbf{i},\mathbf{j}} \equiv \sum^{P}_{p=1} n_{\mathbf{j}_p,\mathbf{j}_{p+1}},
\end{equation}
is extensive, such that for any bipartition of the region in two sub-regions $A_1$, $A_2$ $M(A) = M(A_1) + M(A_2)$.
\begin{figure}
    \centering
    \includegraphics[width=0.5\linewidth]{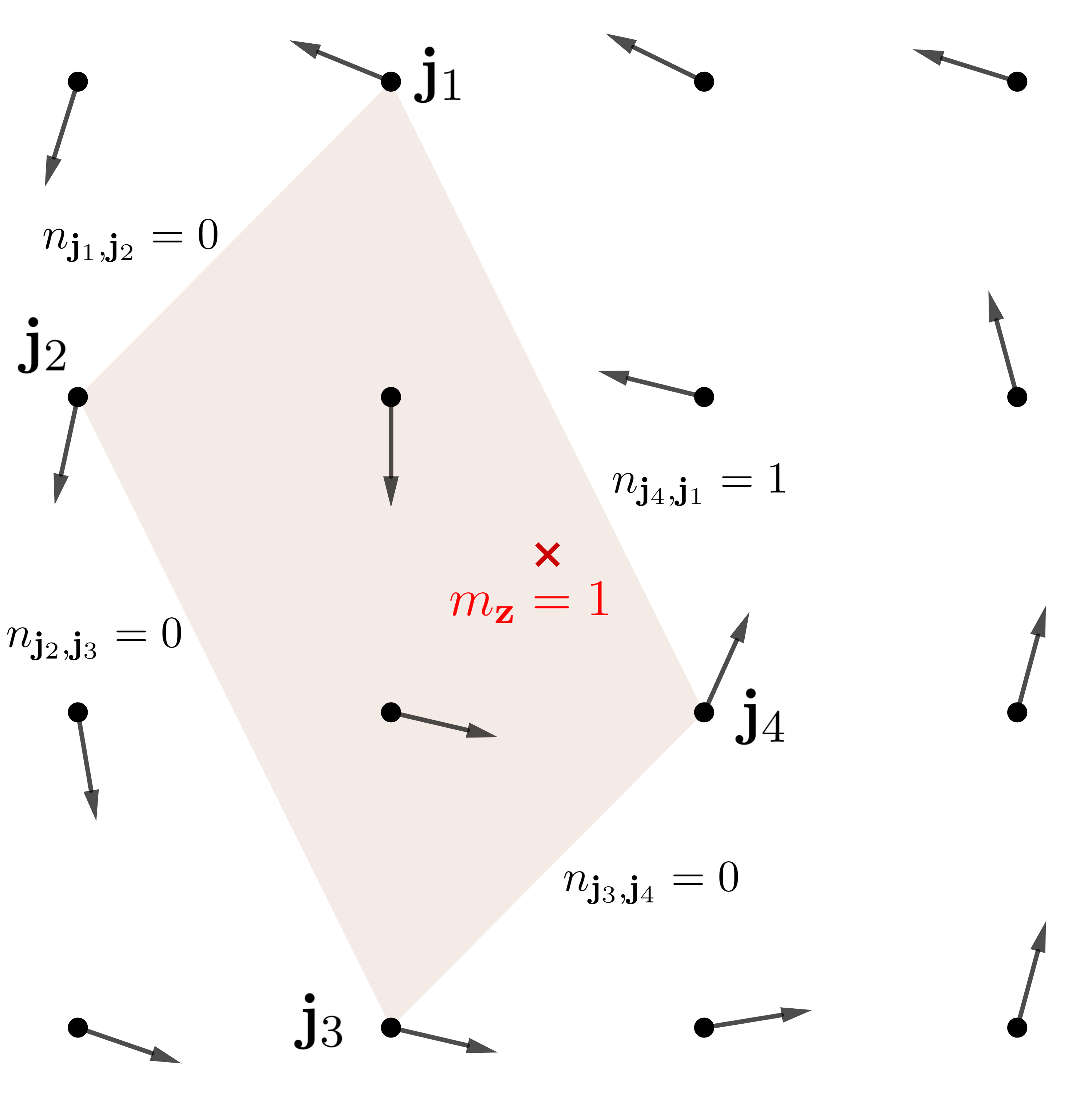}
    \caption{A topological configuration, namely a vortex with charge $m_\mathbf{z} = 1$, and a corresponding choice of the Villain link variables which encircle a parallelogram in such a way to satisfy the constraint \eqref{Villconstr}. In agreement with the symmetry \eqref{gaugesym} the choice of the $n_{\mathbf{i},\mathbf{j}}$ which is different from zero is arbitrary.}
    \label{FigVill1}
\end{figure}
As a consequence, we can introduce for each point of the dual lattice $\mathbf{z}$ its corresponding vortex charge $m_{\mathbf{z}}$ so that the constraint finally reads as
\begin{equation} \label{Villconstr}
    \sum_{\partial A} n_{\mathbf{i},\mathbf{j}} = \sum_{\mathbf{z} \in A} m_{\mathbf{z}}.
\end{equation}
See Fig.\,\ref{FigVill1} for a graphic interpretation of the constraint \eqref{Villconstr} within a prototypical topological configuration. Strictly speaking, one or more of the $\mathbf{z}$ could lie on the boundary  $\partial A$ of the region $A$. This could, in principle, cause some problems. However, as we are going to see, this boundary issue is not a problem in the thermodynamic limit as only the large loops are going to contribute to the critical behavior. 

Finally, the partition function of the model takes the form
\begin{equation} \label{ZVill}
    Z = \sum_{\lbrace m_i \rbrace} \int \prod_{\mathbf{j}} d \theta_{\mathbf{j}} \ e^{-\beta H_0} \delta \left(\sum_{\partial A} n_{\mathbf{i},\mathbf{j}} = \sum_{\mathbf{z} \in A} m_{\mathbf{z}}\right), 
\end{equation}  
with $H_0$ given by \eqref{VHam} and we assume $J(r)$ to have the form of Eq.\,\eqref{Jr}.

\section{Vortex-vortex interaction}
\label{SecVortexVortex}
\noindent
In this section we derive the effective interaction between topological charges $m_{\mathbf{z}}$ in Eq.\,\eqref{ZVill}. To this aim, we define the Fourier transformed variables
\begin{equation}
\theta_\mathbf{j} = \frac{1}{\sqrt{N}} \sum_\mathbf{q} \theta_\mathbf{q} e^{i\mathbf{q} \cdot \mathbf{j}} \hspace{1cm} n_{\mathbf{j},\mathbf{j+r}} =  \frac{1}{\sqrt{N}} \sum_{\mathbf{q}} n_{\mathbf{q},\mathbf{r}} e^{i\mathbf{q} \cdot \left( \mathbf{j} + \frac{\mathbf{r}}{2} \right)},
\end{equation}
with $\mathbf{q}$ a vector of the reciprocal lattice. The Hamiltonian Eq.\,\eqref{VHam} becomes: 
\begin{equation} \label{provvisorio}
\beta H_0 = \frac{1}{4} \sum_{\mathbf{q},\mathbf{r}} J(r) \left| 2 i \sin \frac{\mathbf{q} \cdot \mathbf{r}}{2} \theta_\mathbf{q} - 2 \pi n_{\mathbf{q},\mathbf{r}} \right|^2. 
\end{equation} 
In order to integrate out the $\theta_{\mathbf{j}}$, it is useful to introduce the new variables
\begin{equation} \label{psitheta}
\psi_\mathbf{q} = \theta_\mathbf{q} + \frac{2 \pi i}{K(\mathbf{q})} \sum_\mathbf{r} J(r) \sin \frac{\mathbf{q} \cdot \mathbf{r}}{2} n_{\mathbf{q},\mathbf{r}}
\end{equation}
where
\begin{equation} \label{Kq3}
K(\mathbf{q}) = \sum_\mathbf{r} J(r) \left( 1 - \cos \mathbf{q} \cdot \mathbf{r} \right) = 2 \sum_\mathbf{r} J(r) \sin^2 \frac{\mathbf{q} \cdot \mathbf{r}}{2} 
\end{equation}
In terms of the new variables the Hamiltonian decouples into two pieces, a spin-wave term and a topological term, which only depends on the discrete link variables
\begin{equation}
\beta H_0 = H_{\rm SW}(\psi_j) + H_{\rm top} (m_j)
\end{equation}
The two pieces are given by:
\begin{equation}
\begin{split}
H_{\rm SW} &= \frac{1}{2} \sum_\mathbf{q} K(\mathbf{q}) |\psi_\mathbf{q}|^2 \\
H_{\rm top} &= \pi^2 \sum_\mathbf{q} \frac{1}{K(\mathbf{q})} \sum_{\mathbf{r},\mathbf{r'}} J_r J_{r'} \left|\sin \frac{\mathbf{q} \cdot \mathbf{r}}{2} n_{\mathbf{q},\mathbf{r}} - \sin \frac{\mathbf{q} \cdot \mathbf{r'}}{2} n_{\mathbf{q},\mathbf{r'}} \right|^2.
\end{split}  
\end{equation}
We notice that the spin-wave part of the Hamiltonian exactly corresponds to the quadratic approximation of the original long-range $XY$ Hamiltonian. The asymptotic behavior of $K(\mathbf{q})$ for small $q$ is worked out in Appendix \ref{AppKq}. We find the usual nearest-neighbours scaling relation $K(q) \sim q^2$ for $\sigma > 2$, while for $0 <\sigma < 2$ we have
\begin{equation}
    K(q) = J c_{\sigma} q^{\sigma} + O(q^2)
\end{equation}
with $c_{\sigma}$ an universal function of $\sigma$.

Let us now focus on $H_{\rm top}$: we can notice that, back to the real space, this takes the form
\begin{equation}
H_{\rm top} = \sum_{\mathbf{r},\mathbf{r'}} J_r J_{r'} \sum_{\mathbf{j},\mathbf{j'}} f(\mathbf{j}-\mathbf{j'}) (\nabla_r n_{\mathbf{j},\mathbf{j+r'}}-\nabla_{\mathbf{r'}} n_{\mathbf{j},\mathbf{j+r}}) (\nabla_\mathbf{r} n_{\mathbf{j'},\mathbf{j'+r'}}-\nabla_{\mathbf{r'}} n_{\mathbf{j'},\mathbf{j'+r}}),  
\end{equation}
where 
\begin{equation} \label{fx}
f(\mathbf{x}) = \frac{\pi^2}{4N} \sum_\mathbf{q} \frac{e^{i\mathbf{q} \cdot \mathbf{x}}}{K(\mathbf{q})}
\end{equation}
and we introduced the notation of the discrete curl
\begin{equation} \label{Htoprs}
\nabla_r n_{\mathbf{j},\mathbf{j+r'}}-\nabla_{\mathbf{r'}} n_{\mathbf{j},\mathbf{j+r}} \equiv n_{\mathbf{j},\mathbf{j+r}} + n_{\mathbf{j+r},\mathbf{j+r+r'}} + n_{\mathbf{j+r+r'},\mathbf{j+r}}. 
\end{equation}
As expected, once the continuous field has been integrated out, the interaction depends only on gauge invariants. Indeed, we have that: 
\begin{equation}
    \nabla_r n_{\mathbf{j},\mathbf{j+r'}}-\nabla_{\mathbf{r'}} n_{\mathbf{j},\mathbf{j+r}} = \sum_{\partial \mathcal{P}(\mathbf{j}, \mathbf{r}, \mathbf{r}^{\prime})} n_{\mathbf{i}, \mathbf{i}^{\prime}},
\end{equation}
where the $\mathcal{P} (\mathbf{j}, \mathbf{r}, \mathbf{r}^{\prime})$ denotes the region of the plane corresponding to a parallelogram with vertices  $\mathbf{j}$, $\mathbf{j+r}$, $\mathbf{j+r+r'}$, $\mathbf{j+r}$ (and thus sides $\mathbf{r}$,$\mathbf{r'}$). Thus, as a consequence, the symmetry of Eq.\, \eqref{gaugesym} leaves these quantities invariant.  

By means of the constraint \eqref{Villconstr} finally, we can express the Hamiltonian in terms of the topological charges enclosed in each parallelogram, obtaining an interaction Hamiltonian of the form
\begin{equation} \label{vortexprovvisorio}
H_{\rm top} = \sum_{\mathbf{z},\mathbf{z'}} m_\mathbf{z} m_{\mathbf{z'}} U(\mathbf{z}-\mathbf{z'}), 
\end{equation}
for some potential $U(R)$. The exact form of $U$, along with its large-distance behavior, are derived in Appendix \ref{AppU}. In particular, we find that the small $q$ behavior of the Fourier transform of the potential $U(R)$ is given by
\begin{equation}
  U(q) = \frac{
    2 \pi^2}{\sigma - 1}  \frac{K(q)}{q^4}, 
\end{equation}
We see then the proportionality constant $I(\sigma)$ converges as long as $\sigma >1$. The proportionality constant diverges for $\sigma \rightarrow 1^{+}$ (as seen in Appendix \ref{AppU}, this can be seen as an infrared divergence). This means that, as $\sigma \rightarrow 1^{+}$, in the thermodynamic limit, we cannot excite the vortices for any temperature, so that only the spin wave part $H_{\rm SW}$ of the Hamiltonian survives.

For $1 < \sigma < 2$ the vortex-vortex potential should be taken into account. In Fourier space, for $q \ll 1$, it takes the form: 
\begin{equation} \label{Uq}
U(q) \propto \frac{K(q)}{q^4} \sim J q^{\sigma - 4} + O(q^{-2})
\end{equation}
In turn, this implies that for $R \gg 1$ ($R$ being the distance between a paof vortices)
\begin{equation}
U(R) \sim J \int^{1/R}_{1/L} q^{\sigma-3} dq = \frac{J}{2-\sigma} \left( L^{2 - \sigma} - R^{2 - \sigma} \right)
\end{equation}
The additive constants $L^{2- \sigma}$ brings a term $\propto L^{2-\sigma} \left(\sum_i m_i \right)^2$ in the Hamiltonian \eqref{vortexprovvisorio}, which kills all the non neutral configurations, exactly as in the short-range case. Then, 
\begin{equation}
\begin{split}
H_{\rm top} &= \sum_{i,j} m_i m_j U(\mathbf{r}_i - \mathbf{r}_j) \\
&= U(0) \sum_{i} m_i^2 + \sum_{i \neq j} m_i m_j U(\mathbf{r}_i - \mathbf{r}_j) \\
&= \sum_{i \neq j} m_i m_j \left( U(\mathbf{r}_i-\mathbf{r}_j) - U(0) \right) 
\end{split}
\end{equation}
where now we denoted with $\lbrace \mathbf{r}_i \rbrace$ and $\lbrace m_i \rbrace$ respectively the position and the charge of the vortices of a given configuration, and noticed that, for any neutral configuration $\sum_{i \neq j} m_i m_j = - \sum_i m_i^2 $. We can thus write
\begin{equation} \label{Ur}
U(R) - U(0) \sim J \int \frac{d^2 \mathbf{q}}{(2 \pi)^2} \frac{e^{i \mathbf{q} \cdot \mathbf{R}}-1}{q^{4-\sigma}} \sim - J \ R^{2-\sigma}.
\end{equation}
Let us notice that, now that the sum runs only on $i \neq j$, $U(r)$ is no longer defined up to an additive constant. It is customary, however, to define the energy so that it is zero for $r=1$, which is the minimum distance at which a pacan be created. This means that in the Hamiltonian we can write: 
\begin{equation}
H_{\rm top} = \sum_{i \neq j} m_i m_j \left( U(\mathbf{r}_i-\mathbf{r}_j) - U(1) \right) + \left( U(0) - U(1) \right)  \sum_{i} m_i^2 .
\end{equation}
Now we can finally introduce
\begin{equation} \label{Vvort}
V(\mathbf{r}) = U(\mathbf{r}) - U(1) = g \int \frac{d^2 \mathbf{q}}{2 \pi} \frac{e^{i\mathbf{q} \cdot \mathbf{r}} - e^{i\mathbf{q} \cdot \mathbf{n}}}{q^{4 - \sigma}} = - \gamma \left( r^{2-\sigma} - 1 \right)
\end{equation}
(with $g, \gamma \propto J$) and the finite quantity $\varepsilon_c = U(0) - U(1) >0$, which physically represents the core energy associated with the creation of a vortex paat the minimum possible distance. In terms of these quantities the Hamiltonian becomes
\begin{equation}
H_{\rm top} = \sum_{i \neq j} m_i m_j V(\mathbf{r}_i-\mathbf{r}_j) +  \varepsilon_c \sum_{i} m_i^2
\end{equation}
while, the (topological part of the) partition function is given by
\begin{equation}
    Z_{\rm top} = \sum_{\lbrace m_j \rbrace} \int \prod_j d^2 \mathbf{r}_j e^{- H_{\rm top}}. 
\end{equation}
Finally, in the $\sigma >2$ case we recover the usual logarithmic interaction of the two-dimensional Coulomb gas as
\begin{equation}
U(R) \sim J \int^{1/R}_{1/L} q^{-1} dq = J (\ln L - \ln r).
\end{equation}
Since in this case, as noticed, the dispersion relation $K(q)$ of the spin waves has the same form of the nearest-neighbours one, we can safely conclude that the Villain model is in the same universality class of the nearest-neighbours case for all $\sigma > 2$.

\section{Renormalization and Phase-Diagram of the model}
\label{SecRealRG}
\noindent
We are going to carry out the RG procedure in the vortex gas representation. To this extent we then introduce the renormalization parameter $\ell$ such that the effective lattice spacing $a_{\ell}$ is given by $a_{\ell} = e^{\ell}$. Our picture is the following: at scale $\ell$ the vortex-antivortex pairs with scale $< a_{\ell}$ renormalizes the vortex-vortex potential. While the ultraviolet potential $V_{\ell = 0} (r)$ is given by Eq.\,\eqref{Vvort}, there is no reason for the effective potential at scale $\ell$, $V_{\ell} (r)$, to have the same functional form. 

Since in the ultraviolet the potential grows as a power law a simple energy-entropy scaling argument would suggest that the fugacity $y \equiv e^{- \varepsilon_c}$ is not relevant at any temperature. However, as we are going to see, these naive expectations are defied by the RG calculations, which shows that, for every finite value of $\ell$, the behavior at large distances of the potential is no longer given by the power law of Eq.\,\eqref{Vvort}. The renormalization procedure we present here is the straightforward generalization of those introduced for the short-range case by Kosterlitz and Thouless \cite{KT} to arbitrary (well-behaved) interaction potential and it is similar to those present in \cite{Minnhagen} for a screened Coulomb interaction. However, up to our knowledge, the corresponding treatment for the case of a potential which grows boundlessly is absent in the literature. 

The details of the calculations are given in Appendix \ref{SecDetRG}. The corresponding renormalziation equations for the potential $V_{\ell}$ and the fugacity $y$ turn out to be
\begin{equation} \label{RGVill}
\begin{split} 
\partial_{\ell} V_{\ell} (r) &= r V_{\ell}^{\prime}(r) - V_{\ell}^{\prime} (1) -  \frac{\pi}{2} y_{\ell}^2 \left( \Delta V_{\ell}(r) - \Delta V_{\ell}(1) \right) \\
\frac{d y_{\ell}}{d \ell} &= y_{\ell} \left( 2 + V_{\ell}^{\prime} (1) \right) + O(y^3)
\end{split}
\end{equation}
with $\Delta V$ given by the convolution
\begin{equation} \label{DeltaV}
\Delta V(r) = \int d^2 \mathbf{x} \left( \nabla V_{\ell} (\mathbf{x}) \cdot \nabla V_{\ell} (\mathbf{x} + \mathbf{r}) - |\nabla V_{\ell} (\mathbf{x})|^2 \right).  
\end{equation}

Starting from these equations, we now want to derive the phase diagram of the model, discussing the behavior of the renormalization flow described by Eqs.\,\eqref{RGVill}. Quite surprisingly, in spite of the fact that we are dealing with a non-linear functional set of equations, it is possible to solve them analytically. 

Let us start by noticing that, since $\Delta V_{\ell}$ in Eq.\,\eqref{DeltaV} involves a convolution, it is natural to write $V_{\ell}(r)$ as
\begin{equation}
V_{\ell} (r) = \int \frac{d^2 q}{(2 \pi)^2} U_{\ell} (q) \left( e^{i \mathbf{q} \cdot \mathbf{n}} - e^{i \mathbf{q} \cdot \mathbf{r}} \right), 
\end{equation}
with $\mathbf{n}^2 = 1$ and $U_{\ell = 0} (q) = - g q^{-4-\sigma}$, in agreement with Eq.\,\eqref{Vvort}. In terms of $U_{\ell} (q)$, the first equation of Eq.\,\eqref{RGVill} becomes
\begin{equation}
\partial_{\ell} U_{\ell} (q) = - \left( 2 + q \partial_q \right) U_{\ell} (q) +  \frac{\pi}{2} y_{\ell}^2 q^2 U_{\ell} (q)^2,
\end{equation} 
which, in turn, can be rewritten as
\begin{equation} \label{functional}
\partial_{\ell} U^{-1}_{\ell} (q) = \left( 2  - q \partial_q \right) U^{-1}_{\ell} (q) - \frac{\pi}{2} y_{\ell}^2 q^2. 
\end{equation}
Taking into account our initial condition we can solve this equation by using the ansatz: 
\begin{equation}
U^{-1}_{\ell} (q) = - A_{\ell} q^{4 - \sigma} - B_{\ell} q^2,
\end{equation}
finding: 
\begin{equation} \label{RGAB}
\begin{split}
\frac{d A_{\ell}}{d \ell}  &= - (2 - \sigma) A_{\ell} \\
\frac{d B_{\ell}}{d \ell}  &= \frac{\pi}{2} y_{\ell}^2 
\end{split}
\end{equation}
along with the initial condition $A_0 = g^{-1}$, $B_0 = 0$. We have then that $A_{\ell} \rightarrow 0$ in the infrared, while $B_0$ grows as long as $y_{\ell} \neq 0$. The vortex-vortex potential becomes then:  
\begin{equation} \label{VRG}
V_{\ell} (r) = - \int \frac{d^2 \mathbf{q}}{(2 \pi)^2} \frac{ e^{i \mathbf{q} \cdot \mathbf{n}} - e^{i \mathbf{q} \cdot \mathbf{r}}}{A_{\ell} q^{4 - \sigma}  +B_{\ell} q^2} = - \int \frac{d q}{2 \pi} \frac{1 - \mathcal{J}_0 (qr)}{A_{\ell} q^{3 - \sigma} + B_{\ell} q},
\end{equation}
$\mathcal{J}_k (qr)$ being the $k$-th order Bessel function of the first kind. By computing $V^{\prime} (r)$ we can derive the equation for $y$ which, together with Eqs.\,\eqref{RGAB}, gives the reduced RG set of equation of the model 
\begin{equation} \label{RGVilly}
    \frac{d y_{\ell}}{d \ell}  = y_{\ell} \left( 2 - \int \frac{dq}{2 \pi} \frac{\mathcal{J}_1 (q)}{B_{\ell} + A_{\ell} q^{2-\sigma}} \right) 
\end{equation}
From Eqs.\,\eqref{RGAB} it follows that for any finite $\ell$ we have $B_{\ell} > 0$, so that the second term in the denominator of Eq.\,\eqref{VRG} dominates for $q \ll 1$ and $V_{\ell} (r) \sim - \ln r$ for $r \gg 1$. As announced then, the infrared behavior of the vortex-vortex potential is, for any finite value of $\ell$, qualitatively different to the ultraviolet behavior. In particular, as in the infrared $A_{\ell} \rightarrow 0$, the RG equations \eqref{RGVill} and \eqref{RGVilly} can be approximated as
\begin{equation} \label{VillBKT}
\begin{split}
\frac{d B_{\ell}}{d \ell}  &= \frac{\pi}{2} y^2_{\ell}\\
\frac{d y_{\ell}}{d \ell}  &= y_{\ell} \left( 2 - \frac{1}{2 \pi B_{\ell}} \right),
\end{split}
\end{equation}
i.e. exactly the form of the BKT RG flow for the short-range $XY$ model. 

As a consequence, we also have here a line of stable fixed points $y=0$, $B < \frac{1}{4 \pi}$, corresponding to a phase in which the vortices are not relevant. At the same time it will exist a transition temperature $T_{BKT}$ such that, for $T > T_{BKT}$, the vortices unbind so that the system flows toward a disordered phase. Also in this case, the correlation length $\xi$ in the disorder phase ($T> T_{BKT}$), is expected to exhibit the usual $BKT$ scaling, namely 
\begin{equation}
    \ln \xi \sim (T - T_{BKT})^{-1/2}. 
\end{equation}

Let us notice, however, how here the low-temperature phase ($T < T_ {BKT}$) does not correspond to a quasi-long-range-ordered phase. Indeed, since the vortices are irrelevant and it is not possible to excite vortex-antivortex pair, the constraint \eqref{Villconstr} tells us that in the infrared the link variables $n_{\mathbf{i},\mathbf{j}}$ are suppressed as well. As a consequence, from Eq.\,\eqref{psitheta} we have that in the infrared we have the identification
\begin{equation}
    \theta(x) \rightarrow \psi(x). 
\end{equation}
In turn, since $H_{\rm SW}$ is a quadratic Gaussian model, with the dispersion relation $K(q) \sim q^{\sigma}$, we have the effective action
\begin{equation}
    S \sim - \frac{1}{2} \int d^2 \mathbf{q} \ q^{\sigma} |\theta(\mathbf{q})|^2.
\end{equation}
This, in turn, implies finite magnetization $m$: indeed, being the measure Gaussian and $\mathbf{s} = (\cos \theta, \sin \theta)$, we have
\begin{equation}
    m^2 = \me{\mathbf{s}^2 (0)} = \me{e^{2 i \theta (0)}} = e^{-2 \me{\theta(0)^2}}, 
\end{equation}
while
\begin{equation}
    \me{\theta(0)^2} \sim \int \frac{d^2 \mathbf{q}}{q^{\sigma}},
\end{equation}
which is finite. The same pheonomenology is present for $\sigma < 1$, for any value of the temperature, as the vortices are never relevant in this regime.

\begin{figure}
    \centering
    \includegraphics[width=0.6\linewidth]{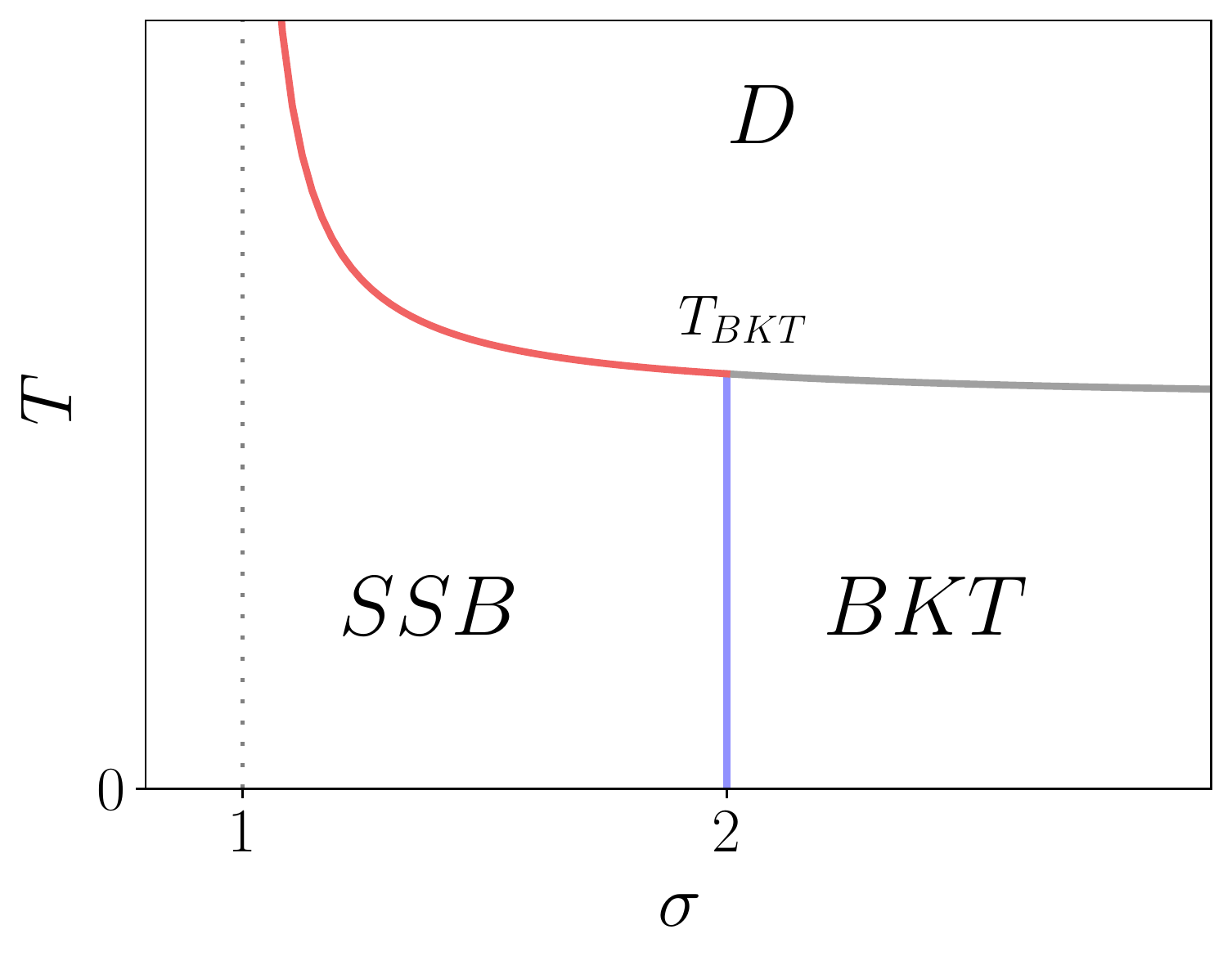}    
    \caption{Qualitative phase diagram of the long-range Villain model. For $\sigma>2$ the system exhibits a BKT transition between the quasi-long-range-ordered BKT phase and the disordered (D) one (grey line). For $1 < \sigma < 2$, at low temperatures the model exhibits a low-temperature spontaneous-symmetry-broken (SSB) phase with a finite magnetization. As the temperature increases, we have a jump in the magenetization to the disordered (D) phase  (red line). This transition, however, still falls within the BKT universality class. For $\sigma \rightarrow 1^{+}$ the transition temperature diverges, so that only the SSB phase survives for $\sigma <1$.}
    \label{FigVill3}
\end{figure}

In summarizing, we found the following phase diagram for the Villain model: 
\begin{itemize}
    \item For $\sigma > 2$, the long-range Villain model exhibits the same phases as its nearest-neighbor counterpart, i.e. it undergoes a BKT phase transition between a low-temperature quasi-long-range ordered phase and a high-temperature disordered one. The phase diagram is thus analogous to the ones of the $XY$ model with $\sigma > 2$. 
    \item For $1 < \sigma < 2$ the model undergoes a phase transition which falls as well under the BKT universality. In this regime, however, we have spontaneous symmetry breaking, and a first-order phase transition in the order parameter as the temperature is increased.
    \item For $0 < \sigma < 1$ the topological defects cannot be excited at any temperature so that the model does not undergo any phase transition and exhibits spontaneous symmetry breaking at any temperature. 
\end{itemize}
A qualitative depiction of the phase diagram of the model is presented in Fig.\,\ref{FigVill3}.
This phase diagram is the main result of our paper. A main comment about it is that it differs
from the corresponding phase diagram of the $XY$ model with long-range interactions for $\sigma<2$,
and therefore we conclude that for $\sigma<2$ the Villain model and the $XY$ model are not longer in the same universality class (notice that our generalization of the Villain model exactly
reduces to the short-range Villain model for $\sigma \to \infty$). This challenges the naive expectation that the long-range Villain model could be mapped in a higher-dimension short-range Villain model as, in $d>2$, the latter is expected to be in the same universality class\,\cite{kleinert1989gauge} of the (short-range) $XY$ model.


\section{Field-theory representation of the model}
\label{SecQFTVill}
\noindent
In this Section we see how some of the results obtained 
through the lattice calculations, starting from the partition function \eqref{ZVill}, can be understood in terms of its continuum limit \eqref{ZVillcont} within the field theory formalism. 

Let us start by expressing the quadratic action in terms of its Fourier modes
\begin{equation}
    \beta H_0 = \frac{J}{4} \int d^2 \mathbf{q} K(q) |\theta (\mathbf{q})|^2 \sim J \int d^2 \mathbf{q} \ q^{\sigma} |\theta (\mathbf{q})|^2, 
\end{equation}
or, in terms of the single-valued, vector field $\mathbf{v} = \nabla \theta (\mathbf{x})$: 
\begin{equation}
    \beta H_0 \sim J \int d^2 \mathbf{q} \ q^{\sigma - 2} |\mathbf{v} (\mathbf{q})|^2.
\end{equation}
It is now possible to solve the constraint in Eq.\,\eqref{ZVillcont} by introducing an auxiliary field $\varphi(\mathbf{x})$ so that
\begin{equation}
Z  \sim  \sum_{\left\lbrace m_i \right\rbrace } \int \mathcal{D} (\mathbf{v}) \int \mathcal{D} (\varphi)  \ e^{- J \int d^2 \mathbf{q} \ q^{\sigma - 2} |\mathbf{v} (\mathbf{q})|^2} \ e^{- i \int d^2 \mathbf{x} \ \varphi(\mathbf{x}) \left( \nabla \times \mathbf{v} - 2 \pi n(\mathbf{x}) \right) }. 
\end{equation}
After noticing that
\begin{equation}
     - i \int d^2 \mathbf{x} \ \varphi (\mathbf{x}) \ \nabla \times \mathbf{v} =  \int d^2 \mathbf{q} \ \mathbf{v} (\mathbf{q}) \times \varphi (\mathbf{q}),
\end{equation}
we can trace out the $\theta$, obtaining, for the topological part of the partition function
\begin{equation} \label{Zintermediate}
Z_{\rm top}  \sim \sum_{\left\lbrace m_i \right\rbrace } \int \mathcal{D} (\varphi)  \ e^{-  J^{-1} \int d^2 \mathbf{q} \ q^{4 - \sigma}   |\varphi (\mathbf{q})|^2 } e^{- 2 \pi i \int d^2 x  \ \varphi(\mathbf{x}) n(\mathbf{x})}
\end{equation}
From here, by integrating out the auxiliary field as well we get 
\begin{equation}
Z_{\rm top}  \sim \sum_{\left\lbrace m_i \right\rbrace } \ e^{- J \int d^2 \mathbf{q} \ q^{\sigma - 4}  |n (\mathbf{q})|^2 } \sim \ \sum_{\left\lbrace m_i \right\rbrace } e^{- \sum_{j,k} m_j m_k U(\mathbf{r}_{j}-\mathbf{r}_{k}) }
\end{equation}
with $U(q) \sim J q^{\sigma-4}$, which is precisely the form of the (large-distance) vortex-vortex interaction we worked out in Sec. \ref{SecVortexVortex}. Let us notice, however, how this field theory approach is not able to predict the divergence in the coupling constant for $\sigma \rightarrow 1$. 

On the other hand, from Eq.\,\eqref{Zintermediate}, one could trace out the vortices, obtaining an effective field theory of the model in terms of the field $\varphi$. To this extent, it is necessary to introduce by hand the core energy of the vortices, whose presence is not captured by the long-distance, field-theoretical description. We get
\begin{equation}
Z_{\rm top}  \sim \sum_{\left\lbrace m_i \right\rbrace } \int \mathcal{D} (\varphi)  \ e^{- J^{-1}  \int d^2 \mathbf{q} \ q^{4 - \sigma}  |\varphi (\mathbf{q})|^2 } e^{- 2 \pi i \int d^2 x  \ \varphi(\mathbf{x}) n(\mathbf{x})} e^{- \varepsilon_c \sum_i m_i^2}.
\end{equation}
Working in the limit of low fugacity $y= e^{-\varepsilon_c}$, we sum over the  configurations such that $m_i = \pm 1$. This means that, for each point in space we get a term of the form 
\begin{equation}
     \sum^{1}_{m=-1} y^m e^{-2 \pi i \varphi(\mathbf{x}) m} = 1 + y \cos 2 \pi \varphi \approx e^{y \cos 2 \pi \varphi }
\end{equation}
obtaining
\begin{equation}
Z_{\rm top}  \sim \int \mathcal{D} (\varphi)  \ e^{-S}, 
\end{equation}
with
\begin{equation} \label{SineGordonStrange}
S = J^{-1} \int d^2 \mathbf{q} \ q^{4-\sigma} |\varphi (\mathbf{q})|^2 - y \int d^2 \mathbf{x} \ \cos(2 \pi \varphi). 
\end{equation}
This is a Sine-Gordon action with a peculiar dispersion relation of the kinetic term $\sim U^{-1} (q)$.  Within this field-theoretical picture, it is the latter that is responsible for the peculiar form of the RG flow. Indeed, $q^{4 - \sigma}$ is less relevant than the usual $q^2$ short-range dispersion relation; on the other hand, if we perform a perturbative RG for small $y$ (e.g. in the Wilson picture) it is known that the Sine-Gordon term, would generate, at the second order in $y$, short-range kinetic terms $\sim q^2 |\varphi(q)|^2$ in the Lagrangian (see e.g. Refs. \cite{gogolin2004bosonization,Giamarchi2004book}). This means that the action \eqref{SineGordonStrange} would flow in the infrared to the usual Sine-Gordon theory, which falls in the universality class of the BKT transition. 

\section{Conclusions}
\noindent
The interplay between topological defects and complex interaction patterns in two- and quasi-two-dimensional systems is known to give rise to complex phase diagrams and exotic scaling behaviors, e.g. in the case of coupled $XY$ planes\,\cite{bighin2019berezinskii}, two-dimensional systems with anisotropic dipolar interactions\,\cite{Maier2004,Fischer2014}, high-dimensional systems with Lifshitz
criticality\,\cite{jacobs1983self,defenu2020topological}, and the anisotropic $3d$ XY model\,\cite{shenoy1995anisotropic}. 

This paper constitutes a further advancement in the understanding of such phenomenology. Here, we studied 
Villain model with a power-law decaying coupling $J(r) \sim J r^{-2 - \sigma}$. Surprisingly, we found that, unlike its $XY$ counterpart, the model falls within the Berezinskii-Kosterlitz-Thouless (BKT) universality class for any $\sigma > 1$. Indeed, while at the lattice level (as expected) the interaction potential between topological defects has a different form in the $\sigma <2$ regime, in the infrared it flows toward the usual logarithmic potential of the two-dimensional Coulomb gas. 

This results 
is noticeably different from the scenario of the long-range $XY$ model, studied in Ref.\,\cite{Giachetti2021} and \cite{Giachetti2022prb}. There, the BKT transition is present only for $\sigma > 7/4$, while an order-disorder transition, described by a new universality class, appears for $\sigma < 2$. 
We can thus conclude 
that the Villain model and the $XY$ model are not in the same universality class for $\sigma < 2$ (see the comparison between Fig.2 of Ref.\,\cite{Giachetti2022prb} and Fig.\ref{FigVill1} of the present paper). This means that, apart from the 
symmetries, in the long-range case also the form of the interaction plays a crucial role. This can be understood by thinking, that even for a smooth configuration, the quadratic approximation of the cosine is not justified, as the interaction between pairs of lattice sites which are far away (and thus uncorrelated) is no longer negligible. As a consequence, the interaction between spin-waves and vortices is not guaranteed to be irrelevant. 

Recent numerical results (see Ref.\,\cite{leuzzi2013,cescatti2019analysis}), suggest that the diluted version of the long-range two-dimensional $XY$ model, in which the lattice sites interact with a probability $\sim r^{-2-\sigma}$, does not reproduce the phase diagram of the long-range $XY$ as well. This suggests, again, that the universality class of non-local $O(2)$ symmetric models in two-dimension is 
rather sensitive to changes in the interaction. 

We conclude that, while in the $1 < \sigma < 2$ region the renormalization-group flow is still described by BKT-like equations, due to the different dispersion relation of the spin-waves the phase diagram (Fig.\,\eqref{FigVill3}) of the long-range Villain model is different from the one of the nearest-neighbor $XY$ model. In particular, the long-range Villain model exhibits a low-temperature ordered phase with a finite magnetization, which jumps discontinuously to zero at $T=T_{BKT}$. This behavior is indeed completely analogous to the one of the one-dimensional Ising model with $1/r^2$ interactions ($\sigma = 1$) \cite{cardy1981one,Froehlich1982,anderson1970exact}. Furthermore, in both systems, the correlation length exhibits the usual BKT scaling while approaching the transition temperature from above.  
Further analytical and numerical investigation is needed in order
to understand whether there is a deeper reason for this and whether this equivalence could be seen already at the lattice level. 


\section*{Acknowledgements}
\noindent This work is supported by the Deutsche Forschungsgemeinschaft (DFG, German Research Foundation) 
under Germany’s Excellence Strategy EXC2181/1-390900948 (the Heidelberg STRUCTURES Excellence Cluster). The 
work is part of the MIUR-PRIN2017 project ``Coarse-grained description for nonequilibrium systems and transport 
phenomena (CO-NEST)'' No. 201798CZL. The authors acknowledge the MISTI Global Seed Funds MIT-FVG collaboration grants ``NV centers for the test of the Quantum Jarzynski
Equality (NVQJE)” and ``Non-Equilibrium Thermodynamics of Dissipative Quantum Systems”.

\section*{Appendices}
\appendix

\section{Behavior of \texorpdfstring{$K(q)$}{Kq} }
\label{AppKq}
\noindent
We will now derive the asymptotic form of $K(q)$. We write down $J(r) = J^S(r) + J r^{-2-\sigma}$ where $J^S(r)$ is a non-universal short-range term. We have then, from  Eq.\,\eqref{Kq3} 
\begin{equation} \label{sr+lr}
K(\mathbf{q}) = 2 \sum_\mathbf{r} J^S(r) \sin^2 \frac{\mathbf{q} \cdot \mathbf{r} }{2} + 2 J \sum_\mathbf{r} r^{-2-\sigma} \sin^2 \frac{\mathbf{q} \cdot \mathbf{r}}{2}. 
\end{equation}
Since, by hypothesis,  $\int_{r>a} d^2  \mathbf{r} \ r^2 J^S (r)$ is finite, we can Taylor expand the cosine in the first integral on the r.h.s getting a term proportional to $q^2$ for small values of $q$. The same is true for the second terms as well, provided that $\sigma >2$, so that we can conclude that 
\begin{equation}
    K(q) \sim q^2 \hspace{0.8cm} \forall \ \sigma > 2 
\end{equation}
with some non-universal proportionality constant. 

Let us consider now the regime $\sigma \in (0,2)$. To understand the small $\textbf{q}$ behavior of the second term in Eq.\,\eqref{sr+lr}, we can replace the sum with an integral
\begin{equation}
2J \sum_\mathbf{r} r^{-2-\sigma} \sin^2 \frac{\mathbf{q} \cdot \mathbf{r}}{2} \approx 2 J \int_{r>1} \frac{d^2 \mathbf{r}}{r^{2+\sigma}}  \sin^2 \frac{ \mathbf{q} \cdot \mathbf{r}}{2} = 2J \int_1^{\infty} \frac{dr}{r^{1+\sigma}} \int_0^{2 \pi} d \theta \sin^2 \frac{q r \cos \theta}{2},
\end{equation} 
or, in term of $\rho = q r/2 \cos \theta $  
\begin{equation}
K(q) = q^{\sigma} 2^{1-\sigma} J \int_{q|\cos \theta|}^{\infty} \frac{d\rho}{\rho^{1+\sigma}} \sin^2 \rho \int_0^{2 \pi} d \theta |\cos \theta|^{\sigma}.
\end{equation}
Now, if $\sigma <2$ the integral has no ultraviolet divergence and we can extend the integral on $\rho$ to the whole axis, committing an error of order $q^2$. Then, we find
\begin{equation}
K(q) = J c_{\sigma} q^{\sigma} + O(q^2) \hspace{0.8cm} \forall \ \sigma \in (0,2) 
\end{equation}
with: 
\begin{equation}
c_{\sigma} = 2^{1-\sigma} \int_{0}^{\infty} \frac{d\rho}{\rho^{1+\sigma}} \sin^2 \rho \int_0^{2 \pi} d \theta |\cos \theta|^{\sigma} =  \frac{2^{1-\sigma} \pi | \Gamma ( \scriptstyle - \frac{\sigma}{2} \displaystyle ) |}{\sigma \Gamma ( \scriptstyle \frac{\sigma}{2} \displaystyle )}
\end{equation}

\section{Derivation of the vortex-vortex potential}
\label{AppU}
\noindent
We now derive the form of the potential $U$ in Eq.\,\eqref{vortexprovvisorio}. We start from the expression of $H_{\rm top}$ Eq.\,\eqref{Htoprs}, taking into account the constraint \eqref{constraintvort}. We find thus
\begin{equation}
H_{\rm top} = \sum_{\mathbf{r},\mathbf{r'}} J_r J_{r'} \sum_{\mathbf{j},\mathbf{j'}} f(\mathbf{j}-\mathbf{j'}) \left( \sum_{\mathbf{z} \in \mathcal{P}(\mathbf{j},\mathbf{r},\mathbf{r'})} m_\mathbf{z} \right) \left(  \sum_{\mathbf{z'} \in \mathcal{P}(\mathbf{j},\mathbf{r},\mathbf{r'})} m_{\mathbf{z'}} \right).  
\end{equation}
Remarkably, only the congruent parallelograms (which share the same $\mathbf{r}$ and $\mathbf{r'}$) do interact. Reshuffling the sums we finally find the vortex-vortex potential
\begin{equation} 
H_{\rm top} = \sum_{\mathbf{z},\mathbf{z'}} m_\mathbf{z} m_{\mathbf{z'}} U(\mathbf{z}-\mathbf{z'}), 
\end{equation}
with
\begin{equation} \label{Ulattice}
U(\mathbf{z}-\mathbf{z'}) = \sum_{  \mathbf{r},\mathbf{r'}} J_r J_{r'} \sum_{\mathbf{j},\mathbf{j'} | \mathbf{z} \in \mathcal{P}(\mathbf{j},\mathbf{r},\mathbf{r'}) \mathbf{z'} \in \mathcal{P}(\mathbf{j'},\mathbf{r},\mathbf{r'}) }  f(\mathbf{j}-\mathbf{j'}).
\end{equation}

\begin{figure}
    \centering
    \includegraphics[width=0.5\linewidth]{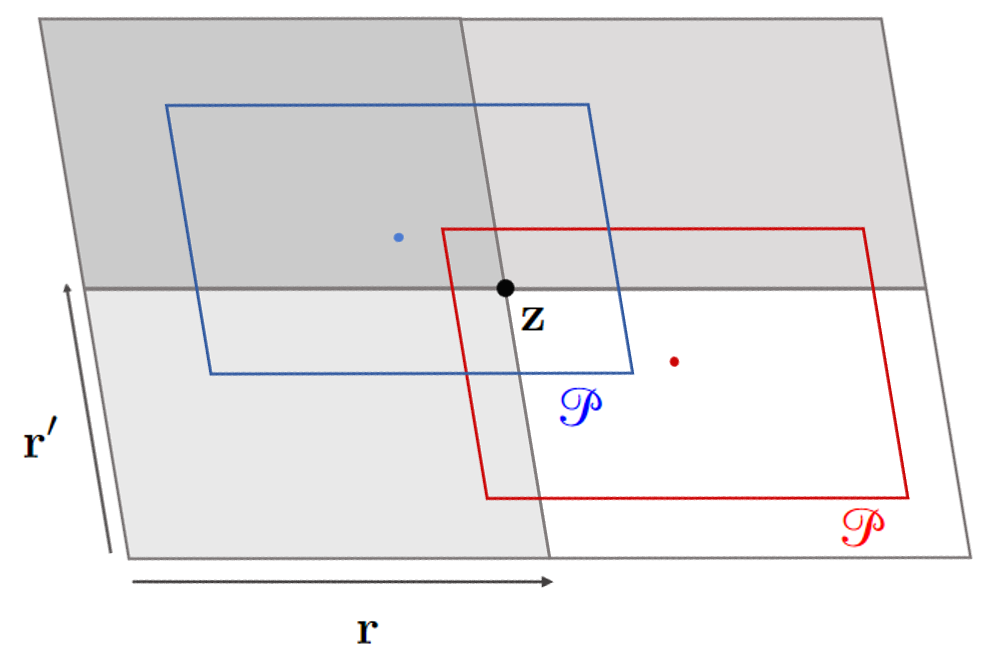}    
    \caption{For a given point of the plane $\textbf{z}$, all the parallelograms $\mathcal{P}$ of sides $\mathbf{r},\mathbf{r'}$ such that $\mathbf{z} \in \mathcal{P}$ (as the red and the blue one in Figure) are such that thecenters belong to a parallelogram of sides $\mathbf{r},\mathbf{r}^{\prime}$, centered in $\mathbf{z}$.}
    \label{FigVill2}
\end{figure}

We want now to derive an approximate expression for this potential valid for large distances. In this limit, we can as well replace the sum in Eq.\,\eqref{Ulattice} with an integral. Since only congruent parallelograms interact we can replace $\mathbf{j}-\mathbf{j'}$ with the distance between the centers of the two. To correctly parametrize the integral, we notice that, once $\textbf{z}$ and fixed $\mathbf{r},\mathbf{r'}$ have been fixed, the parallelogram such that $\mathbf{z} \in \mathcal{P}$ are those whose centers belong to a second parallelogram centered in $\mathbf{z}$ of sides $\mathbf{r}$, $\mathbf{r'}$ (see Fig\,\ref{FigVill2}). 
In turn, each point of this parallelogram can be written as $\mathbf{j} = \mathbf{z} + \lambda \mathbf{r} + \mu \mathbf{r'}$ with $\lambda, \mu \in [-1/2,1/2]$. By choosing $\lambda$, $\mu$ as our coordinates we have to carry a factor $|\mathbf{r} \times \mathbf{r}'|$ due to the jacobian. Finally, we can express $U$ as \begin{equation}
\begin{split}
U(\mathbf{z}- \mathbf{z'}) = &\int d^2 \mathbf{r} \int d^2 \mathbf{r}' (\mathbf{r} \times \mathbf{r}')^2 J(r) J(r') \\ &\int^{1/2}_{-1/2} d \lambda d \lambda' \int^{1/2}_{-1/2} d \mu d \mu' f \left( \mathbf{z}- \mathbf{z'} + (\lambda - \lambda')\mathbf{r} + (\mu - \mu')\mathbf{r}' \right).
\end{split}
\end{equation}
So far, all our considerations are equally valid for any choice of $J(r)$. In what follows we are going to replace $J(r)$ with its asymptotic form $J(r) \sim J r^{-2-\sigma}$ valid for $r \gg 1$. This is safe, since we are interested in the large-distance limit. 

By exploiting the definition \eqref{fx} of $f(\mathbf{x})$, which in the continuum limit becomes
\begin{equation} 
f(\mathbf{x}) =  \frac{\pi^2}{4} \int \frac{d^2 \mathbf{q}}{(2 \pi)^2} \ \frac{e^{i\mathbf{q} \cdot \mathbf{x}}}{K(\mathbf{q})},
\end{equation}
we find the expression for the Fourier transform of $U(\textbf{q})$ of $U(\mathbf{z}-\mathbf{z'})$ to be
\begin{equation}
\begin{split}
U(\mathbf{q}) &= \frac{\pi^2 J^2}{4 K(q)} \int \frac{d^2 \mathbf{r}}{r^{2+\sigma}} \int \frac{d^2 \mathbf{r'}}{r^{' 2+\sigma}} ( \mathbf{r} \times \mathbf{r'} )^2 \int^{1/2}_{-1/2} d \lambda d \lambda' \int^{1/2}_{-1/2} d \mu d \mu' e^{i \mathbf{q} \cdot \left((\lambda - \lambda')\mathbf{r} + (\mu - \mu')\mathbf{r'} \right)} \\
&= \frac{(2\pi J)^2}{K(q)} \int \frac{d^2 \mathbf{r}}{r^{2+\sigma}} \int \frac{d^2 \mathbf{r'}}{r^{' 2+\sigma}} ( \mathbf{r} \times \mathbf{r'} )^2 \frac{\sin^2 (\mathbf{q} \cdot \mathbf{r}/2)}{(\mathbf{q} \cdot \mathbf{r})^2} \frac{\sin^2 (\mathbf{q} \cdot \mathbf{r'}/2)}{(\mathbf{q} \cdot \mathbf{r'})^2} \\
&= \frac{(2\pi J)^2}{K(q)} \int^{2 \pi}_0 d \theta d \theta' \int \frac{dr}{r^{1+\sigma}} \int \frac{dr'}{r^{' 1+\sigma}} \sin^2 (\theta - \theta') \frac{\sin^2 (qr/2 \cos \theta)}{(q \cos \theta)^2} \frac{\sin^2 (qr'/2 \cos \theta')}{(q \cos \theta')^2}.
\end{split}
\end{equation}
Now we notice that, $\forall \sigma <2$, through the substitution $\rho = qr/2 \cos \theta$ 
\begin{equation}
J \int \frac{dr}{r^{1+ \sigma}} \sin^2 (qr/2 \cos \theta) = J 2^{-\sigma} q^{\sigma} |\cos \theta|^\sigma \int \frac{d \rho}{\rho^{1+\sigma}} \sin^2 \rho 
\end{equation}
In Appendix \ref{AppKq} we derived the expression of $K(q)$ for small values $q$, namely $K(q) \sim J c_{\sigma} q^{\sigma}$. By exploiting the expression of $c_{\sigma}$ derived there we have that, for small $q$,
\begin{equation}
  J  \int \frac{dr}{r^{1+ \sigma}} \sin^2 (qr/2 \cos \theta) \sim \frac{1}{2} K(q) |\cos \theta|^{\sigma} \left( \int d \theta^{\prime} |\cos \theta^{\prime}|^{\sigma} \right)^{-1}.
\end{equation}
Finally we have
\begin{equation}
 U(q) \sim I (\sigma) \frac{K(q)}{q^4},
\end{equation}
where
\begin{equation}
\begin{split}
I(\sigma) &=  \pi^2 \left( \int d \theta'' |\cos \theta''|^{\sigma} \right)^{-2}  \int d \theta d \theta' \sin^2 (\theta - \theta') |\cos \theta|^{\sigma-2} |\cos \theta'|^{\sigma-2} \\
&=  2 \pi^2  \frac{\int d \theta \sin^2 \theta |\cos \theta|^{\sigma-2}}{\int d \theta' |\cos \theta'|^{\sigma}} = \frac{2 \pi^2}{\sigma - 1} .
\end{split}
\end{equation}
we find thus the expression we where looking for. If we carefully insert an infrared cutoff $L$, it is easy to see that the divergence for $\sigma \leq 1$ is actually an infrared divergence ($\propto L^{1-\sigma}$, $L$ being the linear size of the system).

\section{Renormalization procedure}
\label{SecDetRG}
\noindent
We now give the details of the derivation of Eq.\,\eqref{RGVill}. In order to carry out the renormalization process, we assume that $y \ll  1$: in this regime only the vortices with charge $\pm 1$ will actually contribute to the renormalization procedure. Then, by the neutrality condition, every possible configuration will have the same number of positive and negative charges, and the partition function takes the form
\begin{equation} \label{Zvortices}
Z_{\rm top} = \sum^{\infty}_{p=0} \frac{y_{\ell}^{2p}}{p!^2} \int_{|\mathbf{r}_i - \mathbf{r}_j|>1} \prod^{p}_{i=1} d\mathbf{r}_i^{+} d^2 \mathbf{r}_i^{-} \ e^{- \sum_{i \neq j} m_i m_j V_{\ell}( \mathbf{r}_i-\mathbf{r}_j)}
\end{equation}
where $\mathbf{r}^{\pm}_i$ are the positions of the vortices with positive/negative charge respectively.  

To perform the renormalization we integrate all over the pasuch that $1 < |\mathbf{r}_i - \mathbf{r}_j| < e^{\delta \ell}$. In order to do so, let us consider the sector with $p$ pairs of vortices in the partition function of Eq.\,\eqref{Zvortices}. There are $p^2$ equivalent ways to choose the pato trace out. Then, assuming the coordinates of this couple to be $\mathbf{r}^{\pm}_p \equiv \mathbf{r}^{\pm}$ we have an additional term given by
\begin{equation*}
\begin{split}
&\frac{y^{2p} p^2}{p!^2} \int_{|\mathbf{r}_i - \mathbf{r}_j|> e^{\delta \ell}} \prod^{p}_{i=1}  d^2 \mathbf{r}^{+}_i  d^2 \mathbf{r}^{-}_i \int_{1<|\mathbf{r}_i - \mathbf{r}_j| < e^{\delta \ell}} d\mathbf{r}^{+}_p  d\mathbf{r}^{-}_p  e^{- \sum^{\prime}_{i \neq j} m_i m_j V (\mathbf{r}_i - \mathbf{r}_j)} \\
&= \frac{y^{2p}}{(p-1)!^2} \int_{|\mathbf{r}_i - \mathbf{r}_j|> e^{\delta \ell}} \prod^{p-1}_{i=1} d\mathbf{r}^{+}_i d\mathbf{r}^{-}_i  e^{- \sum^{\prime}_{i \neq j} m_i m_j V(\mathbf{r}_i - \mathbf{r}_j)} \\ 
&\int_{1<|\mathbf{r}^{+} - \mathbf{r}^{-}| < e^{\delta \ell}} d^2 \mathbf{r}^{+}  d^2 \mathbf{r}^{-} e^{- V(\mathbf{r}^{+} - \mathbf{r}^{-} ) - \sum^{\prime}_i m_i \left( V(\mathbf{r}_i - \mathbf{r}^{+}) - V(\mathbf{r}_i - \mathbf{r}^{-}) \right) },
\end{split}
\end{equation*}
where we are denoting with $\sum^{\prime}$ the summation all over the remaining charges, namely $i,j \neq p$ and we are dropping the subscript $\ell$ in $V(r)$ and $y$ in order to keep the notation easy. This gives nothing but a $p-1$-paterm with the additional interaction 
\begin{equation}
\begin{split}
&\frac{y^{2(p-1)}}{(p-1)!^2} \int_{|\mathbf{r}_i - \mathbf{r}_j|> e^{\delta \ell}} \prod^{p-1}_{i=1}  d^2 \mathbf{r}^{+}_i d^2 \mathbf{r}^{-}_i  e^{- \sum^{\prime}_{i \neq j} m_i m_j V(\mathbf{r}_i - \mathbf{r}_j)} (1 + A) \\ &\approx \frac{y^{2(p-1)}}{(p-1)!^2} \int_{|\mathbf{r}_i - \mathbf{r}_j|> e^{\delta \ell}} \prod^{p-1}_{i=1} d^2 \mathbf{r}^{+}_i  d^2 \mathbf{r}^{-}_i  e^{- \sum^{\prime}_{i \neq j} m_i m_j V(\mathbf{r}_i - \mathbf{r}_j) + \mathcal{A}} 
\end{split}
\end{equation}
where we introduced the quantity
\begin{equation}
\mathcal{A} = y^2 \int_{1<|\mathbf{r}^{+} - \mathbf{r}^{-}| < e^{\delta \ell}} d^1 \mathbf{r}^{+} d^2 \mathbf{r}^{-} e^{- V(\mathbf{r}^{+} - \mathbf{r}^{-} ) - \sum^{\prime}_{i} m_i \left( V(\mathbf{r}_i - \mathbf{r}^{+}) - V(\mathbf{r}_i - \mathbf{r}^{-}) \right) }.
\end{equation}
We now introduce $\boldsymbol{\xi} = \mathbf{r}_{+} - \mathbf{r}_{-}$, $\mathbf{x} = \frac{\mathbf{r}_{+} + \mathbf{r}_{-}}{2}$. Moreover, since $\xi = 1 + O(\delta \ell)$ and $V(1) = 0$, we have that $V(\mathbf{r}^{+} - \mathbf{r}^{-}) = O(\delta \ell^2)$ and can be neglected. At the same time, since the potential is supposed to vary slowly at large distance, we can expand $V(\mathbf{r}_i - \mathbf{r}^{+}) - V(\mathbf{r}_i - \mathbf{r}^{-}) = \boldsymbol{\xi} \cdot \nabla V (\mathbf{r}_i -\mathbf{x}) + O(\xi^3)$. Then we obtain
\begin{equation}
\begin{split}
\mathcal{A} &= y^2 \int_{1< \xi < e^{\delta \ell}} d^2 \boldsymbol{\xi} d^2 \mathbf{x} e^{- \boldsymbol{\xi} \cdot \sum^{\prime}_{i} m_i \nabla V(\mathbf{x}-\mathbf{r}_i)} \\
&= y^2 \int_{1 < \xi < e^{\delta \ell}} d^2 \boldsymbol{\xi} d^2 \mathbf{x}  \left( 1 - \boldsymbol{\xi} \cdot \mathbf{E} + \frac{1}{2} \xi_a \xi_b E_a E_b + O(\xi^4) \right) 
\end{split}
\end{equation}
where we introduced the electric field $\mathbf{E}(\mathbf{x}) = \sum_{i} m_i \nabla V(\mathbf{x}-\mathbf{r}_i)$. Performing the integral over $\boldsymbol{\xi}$: 
\begin{equation}
\mathcal{A} = \text{const} + y^2 (e^{\delta \ell} - 1) \frac{\pi}{2} \int d^2 \mathbf{x} \  \mathbf{E}^2 = \delta \ell  \ \frac{\pi y^2 }{2} \int d^2 \mathbf{x} \ \mathbf{E}^2  + O(\delta \ell^2)
\end{equation}
where we got rid of the additive constant which has no physical meaning. Now we have to compute the electrostatic energy: in the ultraviolet, the electric field of a single charge goes as $\nabla V \sim r^{1-\sigma}$ for $r \gg 1$; taking into account the global neutrality we have $E \sim r^{-\sigma}$ and $E^2 \sim r^{- 2 \sigma}$ so that the integral is convergent only for $\sigma > 1$, which is exactly the range of the parameter we are interested in. In this case we can write
\begin{equation}
\begin{split}
 \int d^2 \mathbf{x} \ \mathbf{E}^2(\mathbf{x}) &= \sum^{\prime}_{i,j} m_i m_j \int  d^2 \mathbf{x} \nabla V(\mathbf{x} + \mathbf{r}_i) \cdot \nabla V(\mathbf{x} + \mathbf{r}_j) \\
&= \sum^{\prime}_{i,j} m_i m_j \int  d^2 \mathbf{x} \nabla V(\mathbf{x}) \cdot \nabla V(\mathbf{x} + \mathbf{r}_j - \mathbf{r}_i) \\
&= \sum^{\prime}_{i \neq j} m_i m_j \int  d^2 \mathbf{x} \left( \nabla V(\mathbf{x}) \cdot \nabla V(\mathbf{x} + \mathbf{r}_j - \mathbf{r}_i) - |\nabla V(\mathbf{x})|^2 \right) ,
\end{split} 
\end{equation}
where we once again used the charge neutrality to write everything in terms of the sum with $i \neq j$. Thus, setting $p^{\prime} = p -1$, we end up with a partition function with the same form of Eq.\,\eqref{Zvortices} with a renormalized two-body interaction given by $V_{\ell} (r) \rightarrow V(r) - \frac{\pi^2}{2} y_{\ell}^2 \Delta V(r)$, with
\begin{equation} \label{DeltaVApp}
\Delta V(r) = \int d^2 \mathbf{x} \left( \nabla V_{\ell} (\mathbf{x}) \cdot \nabla V_{\ell} (\mathbf{x} + \mathbf{r}) - |\nabla V_{\ell} (\mathbf{x})|^2 \right).  
\end{equation}

Finally, in order to correctly write the change of the partition function under the renormalization procedure, we have to redefine the length-scale $\tilde{r} = r e^{-\delta \ell}$ so that the new cutoff length $a_{\ell + \delta \ell}$ is $1$ as well. From the integration measure of Eq.\,\eqref{Zvortices}, we get a factor $e^{4p \delta \ell}$ which can be reabsorbed into a renormalization of the fugacity, $ y_{\ell} \rightarrow y_{\ell} e^{2 \delta \ell }$. In terms of $\tilde{r}$, the potential becomes
$V_{\ell} (r) = V_{\ell} \left( \tilde{r} \right) + \tilde{r} \ V_{\ell}^{\prime} (\tilde{r}) \delta \ell + O(\delta \ell^2)$. From now on we will rename $\tilde{r}$, $r$. Thus, up to higher orders in $\delta \ell$,  we can write $V_{\ell }(r) \rightarrow V_{\ell}(r) + \delta  V(r) $ where
\begin{equation}
\delta V(r) = \delta \ell \left( r V_{\ell}^{\prime} (r) - \frac{\pi y^2}{2} \Delta V(r) \right)
\end{equation}
The new interaction energy, however, no longer respects the condition $V(1) = 0$, so that the procedure cannot be properly repeated. To make up for it we can exploit once again the neutrality condition to have: 
\begin{equation}
\begin{split}
H_{\rm top} &= \sum_{i \neq j} m_i m_j \left( V_{\ell} (\mathbf{r}_i - \mathbf{r}_j) + \delta V (\mathbf{r}_i - \mathbf{r}_j) \right) \\
&= \sum_{i \neq j} m_i m_j  V_{\ell + \delta \ell} (\mathbf{r}_i - \mathbf{r}_j) - \delta V(1)   \sum_{i} m_i^2  
\end{split}
\end{equation}
with $V_{\ell + \delta \ell} (r) = V_{\ell} (r) + \delta V(r) - \delta V(1)$. The last term can be absorbed into the renormalization of the fugacity: 
\begin{equation}
y_{\ell + \delta \ell} =  y_{\ell} e^{ 2 \delta \ell } e^{\delta V(1)}
\end{equation}
In summarizing, we have been able to put the partition function in the same form of the original one, with a different fugacity $y_{\ell + \delta \ell}$ and a different vortex interaction energy $V_{\ell + \delta \ell}(r)$, given by 
\begin{equation} \label{RGVillApp}
\begin{split} 
\partial_{\ell} V_{\ell} (r) &= r V_{\ell}^{\prime}(r) - V_{\ell}^{\prime} (1) -  \frac{\pi}{2} y_{\ell}^2 \left( \Delta V_{\ell}(r) - \Delta V_{\ell}(1) \right) \\
\frac{d y_{\ell}}{d \ell} &= y_{\ell} \left( 2 + V_{\ell}^{\prime} (1) \right) + O(y^3)
\end{split}
\end{equation}
with $\Delta V$ form Eq.\,\eqref{DeltaVApp}. These are the renormalization equations given in the main text.

\bibliography{main}

\end{document}